\documentclass[aps,prl,twocolumn,superscriptaddress,showpacs,preprintnumbers,amsmath,amssymb,showkeys,nofootinbib]{revtex4-2}

\usepackage{graphicx}
\usepackage{dcolumn}
\usepackage{bm}
\usepackage[mathlines]{lineno}
\usepackage[hidelinks]{hyperref}
\usepackage{xcolor}

\hypersetup{
    colorlinks,
    linkcolor={red!50!black},
    citecolor={blue!50!black},
    urlcolor={blue!80!black}
}
\usepackage{multirow}

\def\arrvline{\hfil\kern\arraycolsep\vline\kern-\arraycolsep\hfilneg}

\begin{document}
\newcommand*{\MIT}{Massachusetts Institute of Technology, Cambridge, Massachusetts  02139, USA}
\newcommand*{\MITindex}{1}
\affiliation{\MIT}
\newcommand*{\JLAB}{Thomas Jefferson National Accelerator Facility, Newport News, Virginia 23606, USA}
\newcommand*{\JLABindex}{2}
\affiliation{\JLAB}

\newcommand*{\ANL}{Argonne National Laboratory, Argonne, Illinois 60439, USA}
\newcommand*{\ANLindex}{3}
\affiliation{\ANL}
\newcommand*{\CNU}{Christopher Newport University, Newport News, Virginia 23606, USA}
\newcommand*{\CNUindex}{4}
\affiliation{\CNU}
\newcommand*{\WM}{College of William and Mary, Williamsburg, Virginia 23187, USA}
\newcommand*{\WMindex}{5}
\affiliation{\WM}
\newcommand*{\DUQUESNE}{Duquesne University, Pittsburgh, Pennsylvania 15282, USA}
\newcommand*{\DUQUESNEindex}{6}
\affiliation{\DUQUESNE}
\newcommand*{\FU}{Fairfield University, Fairfield, Connecticut 06824, USA}
\newcommand*{\FUindex}{7}
\affiliation{\FU}
\newcommand*{\FIU}{Florida International University, Miami, Florida 33199, USA}
\newcommand*{\FIUindex}{8}
\affiliation{\FIU}
\newcommand*{\GSIFFN}{GSI Helmholtzzentrum fur Schwerionenforschung GmbH, D-64291 Darmstadt, Germany}
\newcommand*{\GSIFFNindex}{9}
\affiliation{\GSIFFN}
\newcommand*{\JLUGiessen}{II Physikalisches Institut der Universitaet Giessen, 35392 Giessen, Germany}
\newcommand*{\JLUGiessenindex}{10}
\affiliation{\JLUGiessen}
\newcommand*{\INFNFR}{INFN, Laboratori Nazionali di Frascati, 00044 Frascati, Italy}
\newcommand*{\INFNFRindex}{11}
\affiliation{\INFNFR}
\newcommand*{\INFNCAT}{INFN, Sezione di Catania, 95123 Catania, Italy }
\newcommand*{\INFNCATindex}{12}
\affiliation{\INFNCAT}
\newcommand*{\INFNFE}{INFN, Sezione di Ferrara, 44100 Ferrara, Italy}
\newcommand*{\INFNFEindex}{13}
\affiliation{\INFNFE}
\newcommand*{\INFNGE}{INFN, Sezione di Genova, 16146 Genova, Italy}
\newcommand*{\INFNGEindex}{14}
\affiliation{\INFNGE}
\newcommand*{\INFNPAV}{INFN, Sezione di Pavia, 27100 Pavia, Italy}
\newcommand*{\INFNPAVindex}{15}
\affiliation{\INFNPAV}
\newcommand*{\INFNRO}{INFN, Sezione di Roma Tor Vergata, 00133 Rome, Italy}
\newcommand*{\INFNROindex}{16}
\affiliation{\INFNRO}
\newcommand*{\INFNTUR}{INFN, Sezione di Torino, 10125 Torino, Italy}
\newcommand*{\INFNTURindex}{17}
\affiliation{\INFNTUR}
\newcommand*{\SACLAY}{IRFU, CEA, Universit'{e} Paris-Saclay, F-91191 Gif-sur-Yvette, France}
\newcommand*{\SACLAYindex}{18}
\affiliation{\SACLAY}
\newcommand*{\JMU}{James Madison University, Harrisonburg, Virginia 22807}
\newcommand*{\JMUindex}{19}
\affiliation{\JMU}
\newcommand*{\KNU}{Kyungpook National University, Daegu 41566, Republic of Korea}
\newcommand*{\KNUindex}{20}
\affiliation{\KNU}
\newcommand*{\LAMAR}{Lamar University, Beaumont, Texas 77710, USA}
\newcommand*{\LAMARindex}{21}
\affiliation{\LAMAR}
\newcommand*{\MISS}{Mississippi State University, Mississippi State, Mississippi 39762, USA}
\newcommand*{\MISSindex}{22}
\affiliation{\MISS}
\newcommand*{\NMSU}{New Mexico State University, Las Cruces, New Mexico 88003, USA}
\newcommand*{\NMSUindex}{23}
\affiliation{\NMSU}
\newcommand*{\OHIOU}{Ohio University, Athens, Ohio 45701, USA}
\newcommand*{\OHIOUindex}{24}
\affiliation{\OHIOU}
\newcommand*{\ODU}{Old Dominion University, Norfolk, Virginia 23529, USA}
\newcommand*{\ODUindex}{25}
\affiliation{\ODU}
\newcommand*{\SDU}{Shandong University, Qingdao, Shandong 266237, China}
\newcommand*{\SDUindex}{26}
\affiliation{\SDU}
\newcommand*{\MSU}{Skobeltsyn Institute of Nuclear Physics, Lomonosov Moscow State University, 119234 Moscow, Russia}%
\newcommand*{\MSUindex}{27}
\affiliation{\MSU}
\newcommand*{\TEMPLE}{Temple University, Philadelphia, Pennsylvania 19122, USA}
\newcommand*{\TEMPLEindex}{28}
\affiliation{\TEMPLE}
\newcommand*{\GWUI}{The George Washington University, Washington, D.C.~20052, USA}
\newcommand*{\GWUIindex}{29}
\affiliation{\GWUI}
\newcommand*{\ULS}{Universidad de La Serena, Avda Juan Cisternos 1200, La Serena, Chile}
\newcommand*{\ULSindex}{30}
\affiliation{\ULS}
\newcommand*{\UTFSM}{Universidad T\'{e}cnica Federico Santa Mar\'{i}a, Casilla 110-V Valpara\'{i}so, Chile}
\newcommand*{\UTFSMindex}{31}
\affiliation{\UTFSM}
\newcommand*{\INSUBRIA}{Universit\`{a} degli Studi dell'Insubria, 22100 Como, Italy}
\newcommand*{\INSUBRIAindex}{32}
\affiliation{\INSUBRIA}
\newcommand*{\BRESCIA}{Universit\`{a} degli Studi di Brescia, 25123 Brescia, Italy}
\newcommand*{\BRESCIAindex}{33}
\affiliation{\BRESCIA}
\newcommand*{\FERRARAU}{Universit\`{a} di Ferrara , 44121 Ferrara, Italy}
\newcommand*{\FERRARAUindex}{34}
\affiliation{\FERRARAU}
\newcommand*{\Genova}{Universit\`{a} di Genova, 16146 Genova, Italy}
\newcommand*{\Genovaindex}{35}
\affiliation{\Genova}
\newcommand*{\ROMAII}{Universit\`{a} di Roma Tor Vergata, 00133 Rome, Italy}
\newcommand*{\ROMAIIindex}{36}
\affiliation{\ROMAII}
\newcommand*{\ORSAY}{Universit\'{e} Paris-Saclay, CNRS/IN2P3, IJCLab, 91405 Orsay, France}
\newcommand*{\ORSAYindex}{37}
\affiliation{\ORSAY}
\newcommand*{\UCR}{University of California Riverside, Riverside, CA 92521, USA}
\newcommand*{\UCRindex}{38}
\affiliation{\UCR}
\newcommand*{\UCONN}{University of Connecticut, Storrs, Connecticut 06269, USA}
\newcommand*{\UCONNindex}{39}
\affiliation{\UCONN}
\newcommand*{\GLASGOW}{University of Glasgow, Glasgow G12 8QQ, United Kingdom}
\newcommand*{\GLASGOWindex}{40}
\affiliation{\GLASGOW}
\newcommand*{\UNH}{University of New Hampshire, Durham, New Hampshire 03824-3568, USA}
\newcommand*{\UNHindex}{41}
\affiliation{\UNH}
\newcommand*{\SCAROLINA}{University of South Carolina, Columbia, South Carolina 29208, USA}
\newcommand*{\SCAROLINAindex}{42}
\affiliation{\SCAROLINA}
\newcommand*{\VIRGINIA}{University of Virginia, Charlottesville, Virginia 22901, USA}
\newcommand*{\VIRGINIAindex}{43}
\affiliation{\VIRGINIA}
\newcommand*{\YORK}{University of York, York YO10 5DD, United Kingdom}
\newcommand*{\YORKindex}{44}
\affiliation{\YORK}
\newcommand*{\YEREVAN}{Yerevan Physics Institute, 375036 Yerevan, Armenia}
\newcommand*{\YEREVANindex}{45}
\affiliation{\YEREVAN}

\newcommand*{\NOWUCONN}{University of Connecticut, Storrs, Connecticut 06269, USA}
\newcommand*{\NOWWM}{College of William and Mary, Williamsburg, Virginia 23187, USA}
\newcommand*{\NOWTELAVIV}{University of Tel Aviv, Tel Aviv 6997801, Israel}
\newcommand*{\NOWDOTA}{DOTA, ONERA, Universit\'{e} Paris-Saclay, 91120, Palaiseau, France}

\author{S.~Lee}
\email[Contact author: ]{sangbaek.lee@temple.edu}
\altaffiliation[Current address:~]{\TEMPLE}
\affiliation{\MIT}
\affiliation{\ANL}
\author{T.B.~Hayward}
\affiliation{\MIT}
\author{L.~Elouadrhiri}
\affiliation{\JLAB}
\author{R.G.~Milner}
\affiliation{\MIT}

\author {P.~Achenbach} 
\affiliation{\CNU}
\author {J. S. Alvarado} 
\affiliation{\ORSAY}
\author {M.J.~Amaryan}
\affiliation{\ODU}
\author {W.R.~Armstrong} 
\affiliation{\ANL}
\author {H.~Avakian}
\affiliation{\JLAB}
\author {N.A.~Baltzell} 
\affiliation{\JLAB}
% ...

\collaboration{CLAS Collaboration}
\noaffiliation

\title{
Multi-Differential DVCS Cross Section Measurement on the Proton in the Valence Region with CLAS12
}


\date{\today}
             
\pagebreak

\begin{abstract}
We report the measurement of the four-fold differential cross section of deeply virtual Compton scattering (DVCS) on the proton using a longitudinally polarized 10.6~GeV CEBAF electron beam incident on a liquid-hydrogen target with the CLAS12 detector. The analysis provides 1312 data points, expanding the measured phase space in the valence quark region, $0.06 < x_B < 0.58$, $1.00 < Q^2 < 5.76~\mathrm{GeV}^2$, and $0.11 < |t| < 1.00~\mathrm{GeV}^2$.
DVCS cross section measurements directly constrain the real part of the DVCS amplitude through its interference with the Bethe-Heitler process, making them essential inputs to global extractions of the Compton Form Factors (CFFs), particularly the dominant CFF $H$. These data improve sensitivity to these observables and can contribute to future determinations of the proton's internal mechanical structure.
\end{abstract}
\maketitle

Describing the internal structure of hadrons and nuclei in terms of their quark and gluon degrees of freedom is one of the main goals of both theoretical and experimental hadronic physics.
Many fundamental properties such as the internal pressure and shear forces, angular momentum distributions, and their contributions to nucleon spin, remain largely unknown despite decades of study \cite{Achenbach:2023pba}.  

Generalized parton distributions (GPDs) provide a multi-dimensional framework that unifies the longitudinal momentum and transverse spatial distributions of partons inside the nucleon~\cite{PhysRevD.55.7114,https://doi.org/10.1002/prop.2190420202,PhysRevD.56.5524,Ji:1996ek,Ji:1995sv,PhysRevLett.74.1071}. In the Bjorken regime, characterized by large photon virtuality and finite scaling variable, GPDs become experimentally accessible through hard exclusive processes.  

\begin{figure}[!ht]
   \centering
   \includegraphics[width = 0.4\textwidth]{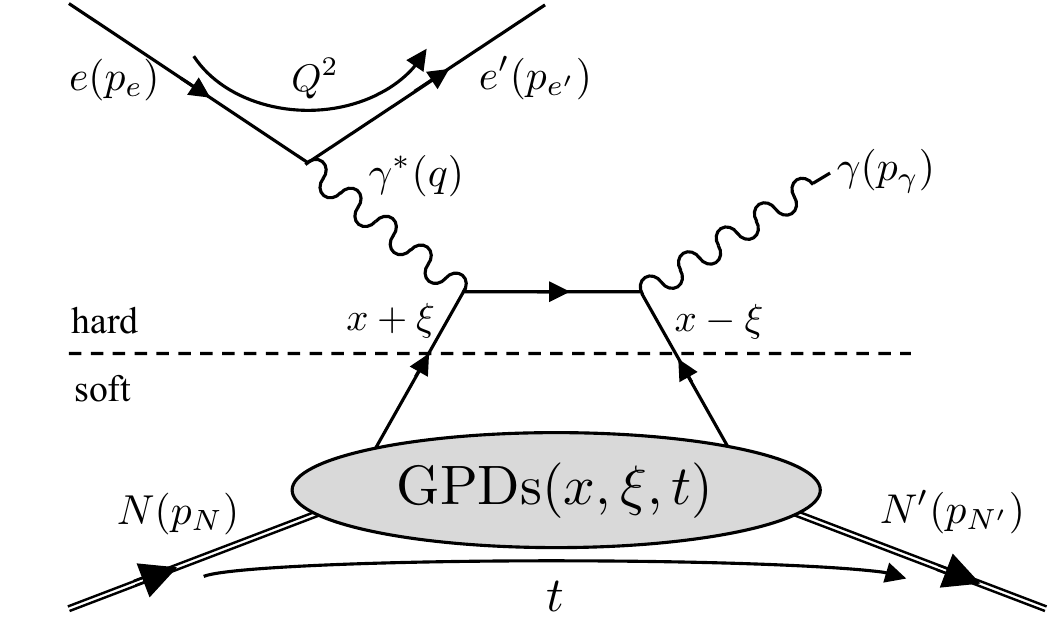}
\caption{The ``handbag'' diagram for the DVCS process on the nucleon $eN \rightarrow e^\prime N^\prime \gamma$. The four-vectors of the incoming/outgoing electrons, photons, and nucleons are denoted by $p_e/p_{e'}$, $q/p_{\gamma}$, and $p_N/p_{N'}$, respectively. Lorentz-invariant variables are illustrated in the figure, including the average partonic light-cone momentum fraction $x$, the skewness $\xi = x_B\left(1+t/Q^2\right)$/$\left(2-x_B\left(1-t/Q^2\right)\right)$, where $x_B$ is the Bjorken-scaling variable $x_B = Q^2/(2p_N \cdot q)$, the four-momentum transfer squared $t~=~(p_N~-~p_{N'})^2$, and the photon virtuality $Q^2 = -(p_{e'} - p_e)^2$.} 
\label{fig:handbag}
\end{figure}

Deeply virtual Compton scattering (DVCS), characterized by the emission of a high-energy photon from the struck parton, provides the cleanest probe of GPDs and enables multi-dimensional imaging of the nucleon~\cite{Burkardt:2000za}. At sufficiently large momentum transfer squared $Q^2$, and fixed Bjorken variable $x_B$, the DVCS amplitude factorizes into a perturbatively calculable hard scattering kernel and a non-perturbative nucleon structure component encoded in the GPDs~\cite{Ji:1998xh, PhysRevD.59.074009,PhysRevD.66.014017}. The DVCS process is commonly represented by the handbag diagram shown in Fig.~\ref{fig:handbag}, in which the hard scattering subprocess factorizes from the soft nucleon structure described by the GPDs.

Experimentally, DVCS interferes coherently with the purely electromagnetic Bethe-Heitler (BH) process, in which a real photon is radiated by the incoming or outgoing lepton. The resulting four-fold differential cross section can be expressed using notations in Refs.~\cite{BELITSKY2002323, Braun:2014sta}

\begin{align}
\frac{d^4\sigma_{ep\rightarrow e'p'\gamma}}{dx_B\, dQ^2\, d|t|\, d\phi}
= 2\pi \Gamma \Big(|\mathcal{T}_{\rm DVCS}|^2 + |\mathcal{T}_{\rm BH}|^2 + \mathcal{I}\Big),
\label{eqn:differential_xsec}
\end{align}

where $\Gamma$ is a kinematic prefactor related to the virtual photon flux, $\mathcal{T}_{\rm DVCS}$ and $\mathcal{T}_{\rm BH}$ are the scattering amplitudes of the DVCS and BH processes, and $\mathcal{I}~=~\mathcal{T}_{\rm DVCS}\mathcal{T}_{\rm BH}^{*} + \mathcal{T}_{\rm BH}\mathcal{T}_{\rm DVCS}^{*}$ is the interference term. In the Trento convention~\cite{Bacchetta:2004jz}, the angle $\phi$ is defined between the lepton scattering plane and the plane spanned by the virtual photon and recoil proton. The pure DVCS and interference contributions to the $ep \rightarrow e'p'\gamma$ cross section can be represented as the sum of a finite number of harmonics of $\phi$ with coefficients that are linear and bilinear combinations of the Compton form factors (CFFs). These CFFs are complex-valued convolutions of the GPDs. The characteristic azimuthal dependences of DVCS observables such as the cross section or spin asymmetries arise from the interference and pure DVCS terms, and provide direct sensitivity to the CFFs \cite{BELITSKY2002323} that are related to the underlying GPDs. 

The GPDs have also been connected to the gravitational form factors (GFFs), which has lead to initial extractions of fundamental mechanical properties of the nucleon, such as internal pressure and shear force distributions~\cite{Burkert:2018bqq, Shanahan:2018nnv, Kumericki:2019ddg}. The $D(t)$ term, or Druck term, probes the pressure and shear forces inside the proton~\cite{Pefkou:2021fni, Duran:2022xag, Burkert:2023wzr, Hackett:2023rif}, and the quark contributions $D^q(t)$ can be extracted from the subtraction term in the dispersion relation~\cite{Anikin:2007tx, Diehl:2007jb} that connects the imaginary and real parts of the CFFs.

The sensitivity of DVCS observables to different CFFs has been studied in Refs.~\cite{Kumericki:2009uq, GuidalM, Kumericki:2016ehc, Shiells:2021xqo}. The complementary sensitivities of different observables to the real and imaginary parts of the CFFs make absolute cross section measurements indispensable. Observables in DVCS include unpolarized and polarized absolute cross section measurements, as well as beam-, target- and double-spin asymmetries. Single-spin asymmetries are primarily sensitive to the interference terms between the BH and DVCS amplitudes and are dominated by the imaginary parts of the CFFs. In contrast, the absolute cross section gives additional access to the squared DVCS amplitude and provides information about both the real and imaginary parts of the CFFs. Measuring the real parts is critical because they enter the dispersion relations that connect the amplitude at a given point to the integral of the GPD over the full momentum range. Precise measurements of the full DVCS cross section in a multi-dimensional space thus contribute to constraining the real parts of the CFFs and enable quantitative extractions of the mechanical properties of the proton.

Previous measurements of the unpolarized DVCS cross section have been reported by several experiments~\cite{JeffersonLabHallA:2006prd, PhysRevLett.115.212003, PhysRevC.92.055202, JeffersonLabHallA:2022pnx}. Early results from Hall A at Jefferson Lab~\cite{JeffersonLabHallA:2006prd, PhysRevC.92.055202} established leading-twist dominance of the CFFs by demonstrating negligible $Q^2$ dependence in the CFFs, consistent with the perturbative QCD scaling behavior of the DVCS cross section~\cite{Ji:1996ek}. However, subsequent analysis~\cite{Defurne2017} of the 5.55~GeV data suggested that higher-twist contributions may not be negligible, motivating further measurements over an extended kinematic range.

\begin{figure}[!ht]
   \centering
   \includegraphics[width = 0.47\textwidth]{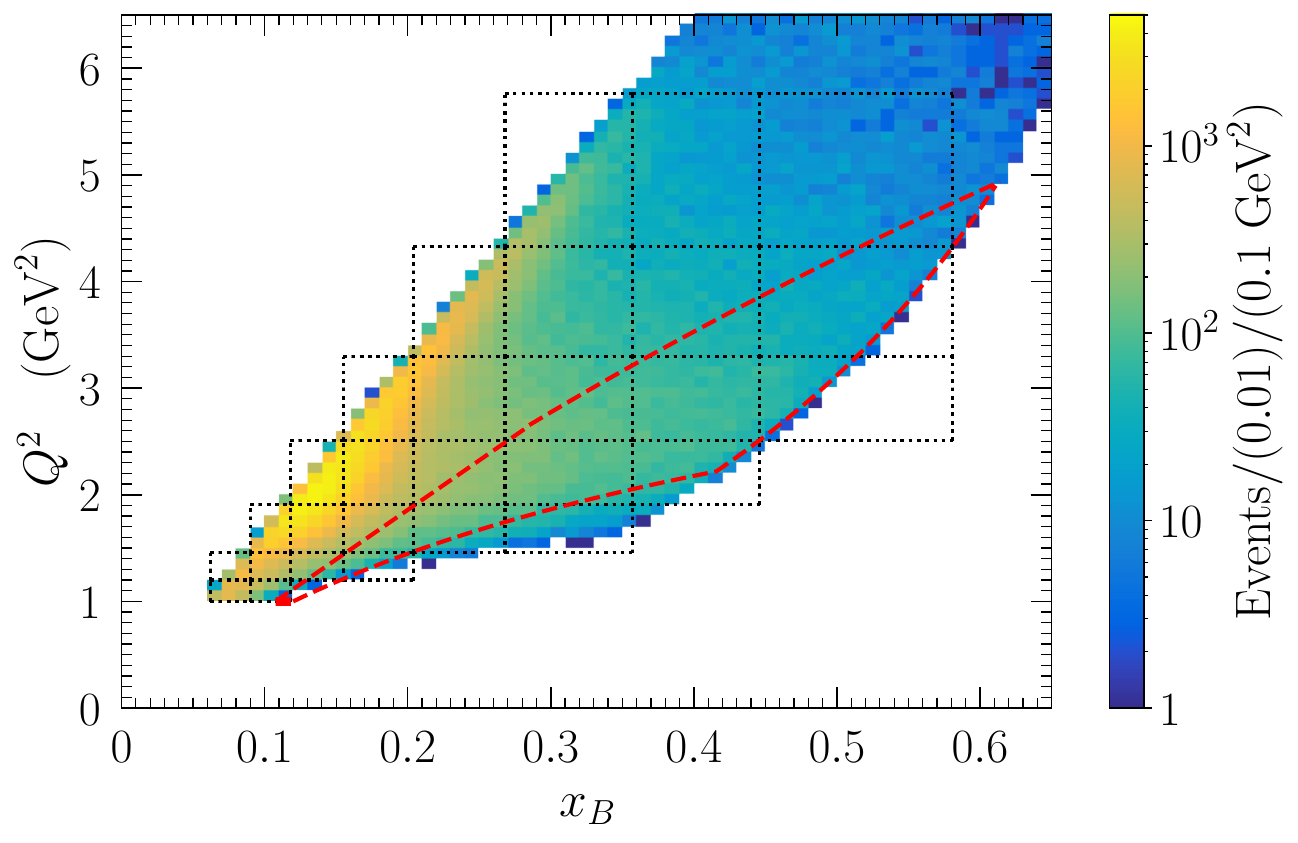}
   \caption{ Virtuality of the
  incoming photon $Q^2$ as a function of the Bjorken variable $x_B$ for selected DVCS events.
   The red dashed curves represent the phase space boundary of the CLAS DVCS experiment at 5.75 GeV \cite{PhysRevLett.115.212003}.}
   \label{fig:binning}
\end{figure}

The present measurement, performed with the CLAS12 detector~\cite{Burkert:2020akg} at 10.6 GeV, expands the accessible phase space in the valence region, covering $0.06<x_B<0.58$, $1.00<Q^2<5.76~\mathrm{GeV}^2$, and $0.11<|t|<1.00~\mathrm{GeV}^2$, and provides 1312 new differential cross section data points. 
The data were collected in fall 2018 using a 10.6~GeV longitudinally polarized electron beam incident on an unpolarized liquid-hydrogen target.
The dataset spans a broad region of phase space as shown in Fig.~\ref{fig:binning} and includes full reconstruction of the final state, $e'$, $p'$, and $\gamma$, in the deep-inelastic regime defined by $W > 2$~GeV, $Q^2 > 1$~GeV$^2$, $E_{e'}>2$~GeV, and $E_{\gamma}>2$~GeV. Here, $W = \sqrt{(q + p_N)^2}$ is the invariant mass of the virtual photon--proton system, and $E_{e'}$ and $E_{\gamma}$ are energies of scattered electron and real photon, respectively.

CLAS12 is a large-acceptance spectrometer located in experimental Hall B at Jefferson Lab \cite{Adderley:2024czm} comprising independent forward and central detector systems that respectively cover polar angles below and above approximately 35 degrees. The forward system consists of drift chambers~\cite{Mestayer:2020saf} within a toroidal magnetic field provided by six symmetrically arranged superconducting coils~\cite{Fair:2020yfx} for the tracking of charged-particles, complemented by Cherenkov counters~\cite{Sharabian:2020whm, Ungaro:2020hbs, Contalbrigo:2020lnd} and electromagnetic calorimeters~\cite{Asryan:2020iqj} for electron identification, and a scintillator time-of-flight system~\cite{Carman:2020fsv} for the identification of charged hadrons. The central system operates within a solenoidal magnetic field~\cite{Fair:2020yfx} and includes silicon strip and Micromegas trackers~\cite{Antonioli:2020ylv, Acker:2020qkv}, together with a time-of-flight system~\cite{Carman:2020yma}. DVCS event candidates were selected from events containing at least one electron, one proton, and one photon. Scattered electrons were detected in the forward detector (FD) system, protons predominantly in the central detector (CD) system, and photons in either the forward calorimeters or the Forward Tagger (FT)~\cite{Acker:2020brf}, which extends coverage to very forward angles of between 2.5 to 4.5~degrees. 

Electrons were identified through the use of the CLAS12 high threshold Cherenkov counter, which provided separation between leptons and hadrons up to 4.9~GeV, in combination with the calorimeter of CLAS12. The same calorimeter system was used to identify photons above 5~degrees with additional supplemental identification in the FT system. Protons were identified in the forward and central time-of-flight systems through comparison of particle velocity and momentum. The exclusivity of the reaction was ensured by applying 3-$\sigma$ cuts on the relevant kinematic variables such as the squared missing mass, missing energy, minimal missing transverse momentum, and the coplanarity angle defined as the difference between computing the angle between the leptonic and hadronic planes using the nucleon and virtual photon versus the virtual and real photons. Contributions from the BH colinear singularity were suppressed by requiring the angle between the scattered electron and the real photon to exceed 8~degrees.

\begin{figure}[!ht]
   \centering
   \includegraphics[width = 0.45\textwidth]{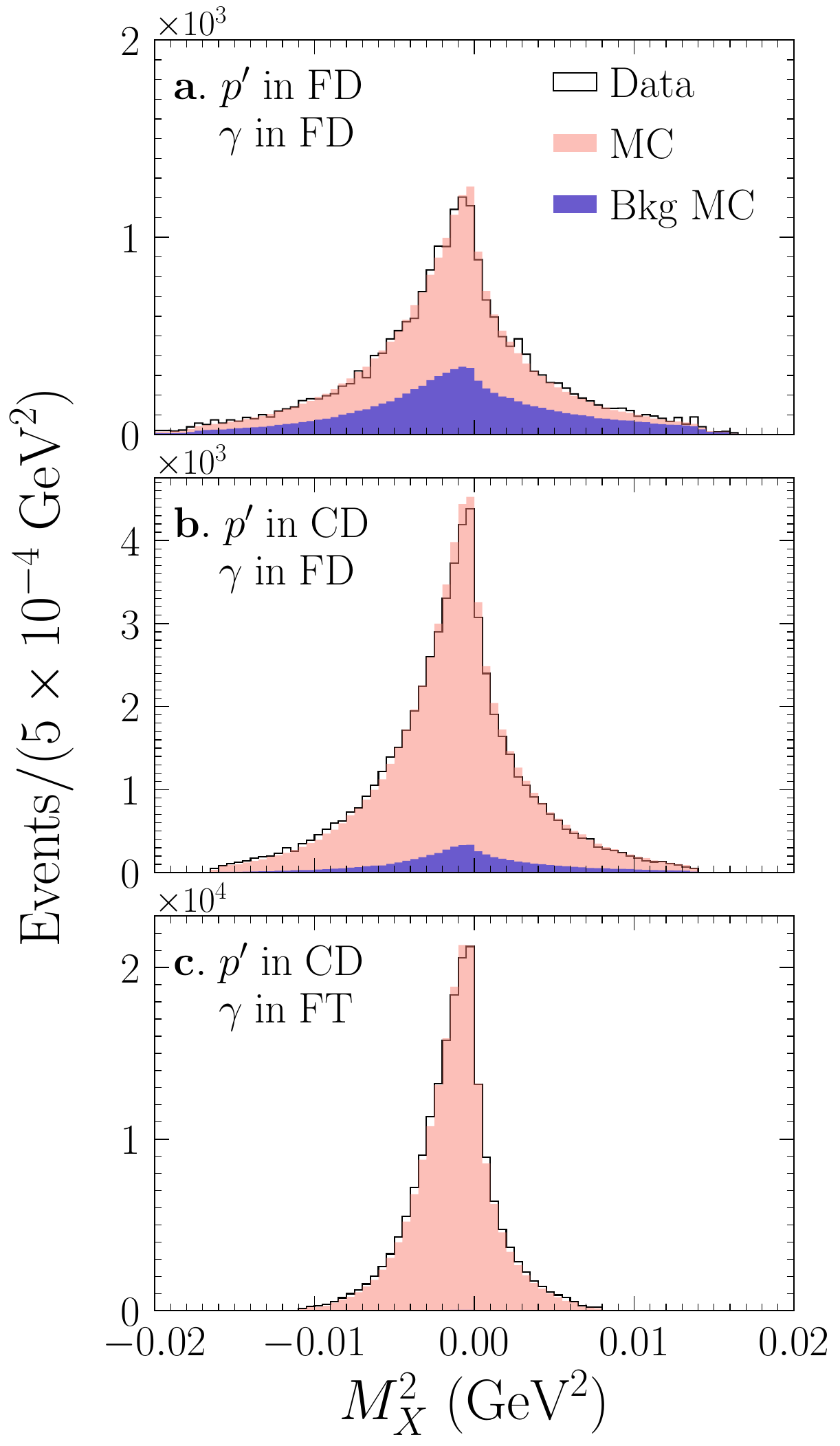}
   \caption{Distributions of $M^2_X(ep\rightarrow e'p'\gamma X)$ for exclusively selected events. Panel (a) shows data (black line), signal-plus-background MC (light red), and background-only MC (dark blue) with both proton and photon in the FD. Panel (b) shows protons in the CD and photons in the FD, while panel (c) shows protons in the CD and photons in the FT. Simulations were normalized to match the data for shape comparison.}
   \label{fig:mm2_dist}
\end{figure}

The $ep \rightarrow e'p'\gamma$ channel receives a significant background from exclusive $\pi^0$ production, where only a single decay photon is detected and the event passes the exclusivity criteria for the single photon process. A dedicated Monte Carlo (MC) sample for the $ep \rightarrow e'p'\pi^0$ process was produced with the CLAS12 simulation~\cite{Ungaro:2020xlc}, generating events weighted by an exclusive $\pi^0$ cross section model based on previous structure-function parameterizations from CLAS measurements in the kinematic region~\cite{CLAS:2012cna, CLAS:2014jpc}. The MC overestimated the experimental detector resolution, hence the electron, proton, and photon kinematics were smeared to reproduce the resolutions observed in data. Simulated yields were scaled to match the measured $ep \rightarrow e'p'\pi^0$ yields as a function of proton kinematics and were used to estimate the residual background in the $ep \rightarrow e'p'\gamma$ sample. Figure~\ref{fig:mm2_dist} shows the total missing mass, $M^2_X(ep \rightarrow e'p'\gamma X)$, distributions for both data and MC across the detector topologies used in this analysis. The large contribution of the $\pi^0$ background occurs when both the proton and photon are detected in the FD, while minimal background is observed when the proton is detected in the CD and the photon in the FT.

The experimental four-fold differential cross section is given by

\begin{align}
\frac{d^4\sigma_{ep\rightarrow e'p'\gamma}}{d x_B\, dQ^2\, d|t|\, d\phi}
=
\frac{
N_{ep\rightarrow e'p'\gamma}
}{
\mathcal{L}\ 
V
A\ \epsilon_{\mathrm{det}}\ F_{\mathrm{bin}}\  F_{\mathrm{rad}}
}, \label{eqn:xsec}
\end{align}
where the raw yields are denoted by $N_{ep\rightarrow e'p'\gamma}$, the integrated luminosity $\mathcal{L}$ is measured with the Hall B Faraday Cup~\cite{Baltzell:2020nvm}, $V$ is the physical phase-space volume, $A$ is the detector acceptance for the $ep~\rightarrow~e'p'\gamma$ channel estimated from the CLAS12 simulation, $\epsilon_{\mathrm{det}}$ is the detector efficiency, $F_{\mathrm{bin}}$ is the bin-centering correction factor, and $F_{\mathrm{rad}}$ is the radiative correction factor. This measurement uses data corresponding to integrated luminosities of 40.1~fb$^{-1}$ and 42.7~fb$^{-1}$ for torus polarities bending electrons towards and away from the beamline, respectively. The detector efficiency, $\epsilon_{\text{det}}$, was determined using CLAS12 Geant4 simulations with random-trigger background merging and included a normalization factor accounting for residual data–MC discrepancies, as described in Supplemental Material S1. The bin-centering factor $F_{\mathrm{bin}}$ was computed as the ratio of the model cross sections averaged over each  $(x_B,Q^2,|t|,\phi)$ bin to the model cross section evaluated at the measured bin-averaged kinematics of $x_B$, $Q^2$, $|t|$, and $\phi$. Radiative corrections were applied by comparing radiative and non-radiative MC datasets produced following the prescriptions in Ref.~\cite{Vanderhaeghen:2000ws}.

Systematic uncertainties include contributions from event selection, fiducial cuts, momentum resolution, background subtraction, acceptance model dependence, radiative corrections, bin-centering corrections, and overall normalization. Bin-dependent uncertainties were evaluated by varying analysis parameters around their nominal values. Acceptance and bin-centering uncertainties were estimated using three cross section models: pure BH, the VGG model~\cite{Vanderhaeghen:1999xj}, and the KM15 model~\cite{Kumericki:2015lhb, Kumericki:2016ehc}, with the standard deviation taken as the systematic uncertainty. The VGG projections were obtained using the \textit{dvcsgen} event generator \cite{dvcsgen} with model parameters constrained by previous measurements \cite{CLAS:2008ahu}, while the KM15 projections were obtained using the \textit{gepard} software \cite{km1}. The dominant contributions are from the fiducial cuts (10.1\%), event selection (8.1\%), 
momentum resolution (7.4\%), and acceptance model dependence (5.8\%), yielding a 
total bin-by-bin systematic uncertainty of 18.6\%.
An alternative radiative correction method~\cite{Akushevich:2017kct} was used to estimate the radiative uncertainty, which was estimated to be below 3\%. Residual data–MC discrepancies were accounted for using a normalization factor derived from $ep\rightarrow e'p'\pi^0$ yields, with a mean value of 1.35, leading to a uncertainty in the normalization factor of 31\%. The total systematic uncertainty of approximately 37\% was obtained by adding the bin-by-bin and normalization uncertainties in quadrature.

\begin{figure}[!ht]
  \includegraphics[width=0.45\textwidth]{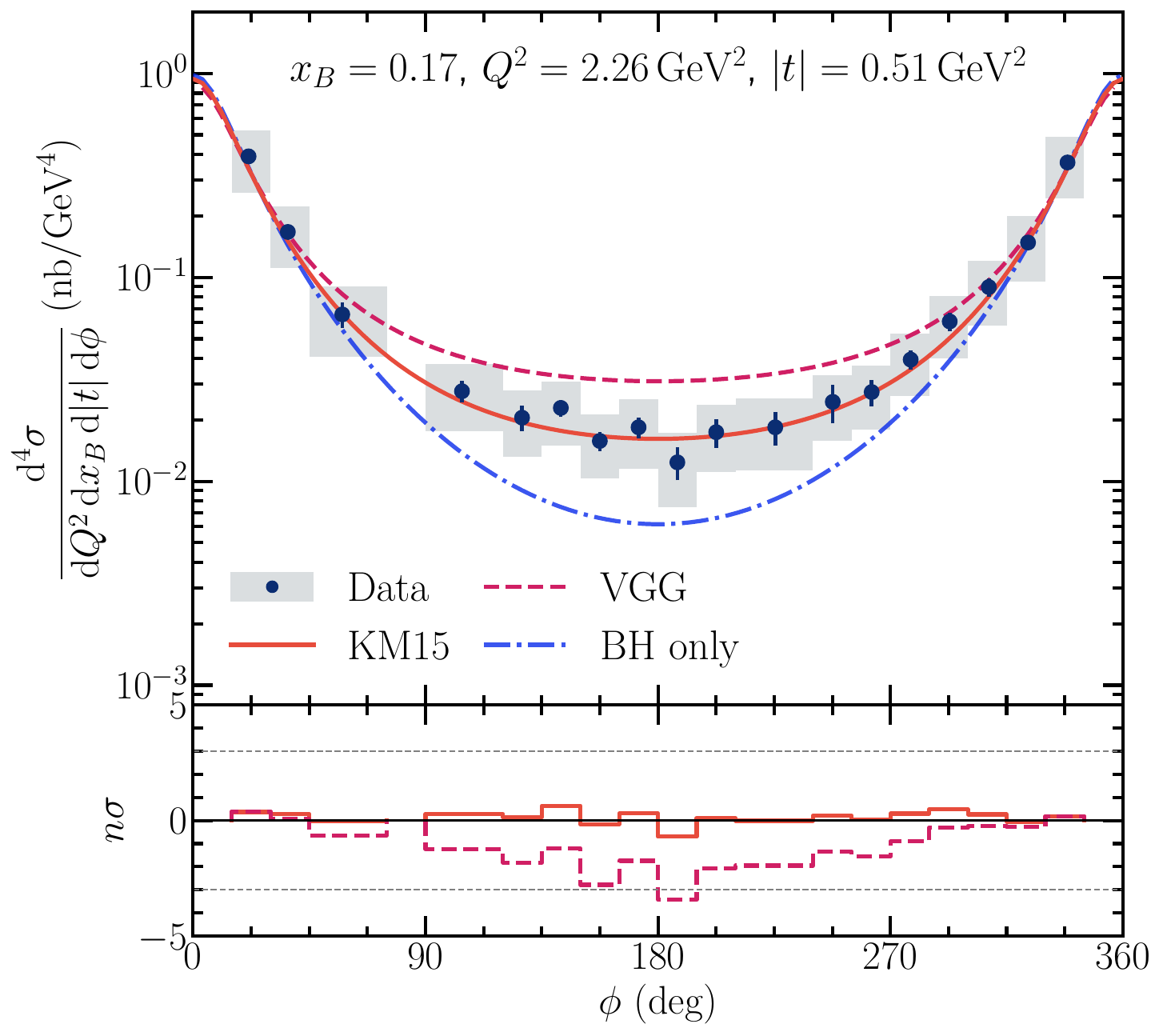}
    \caption{Unpolarized cross sections $\frac{d^4\sigma_{ep\rightarrow e'p'\gamma}}{dx_B dQ^2 d|t| d\phi}$ 
as a function of $\phi$ at $x_B = 0.17$, $Q^2 = 2.26$~GeV$^2$, and $|t| = 0.51$~GeV$^2$. Dark blue points represent the extracted cross sections with statistical (vertical bars) and systematic (gray bands) uncertainties. The blue dash-dotted, red dashed, and orange solid curves show the BH, VGG \cite{Vanderhaeghen:1999xj}, and KM15 \cite{Kumericki:2015lhb, Kumericki:2016ehc} predictions, respectively. The bottom panel shows the residuals between the data and the model predictions normalized to the total uncertainty for VGG and KM15, using the same graphic style. The mean pull with respect to KM15 is 0.15, indicating a consistent description of the data at this kinematics.}
    \label{fig:one_bin}
\end{figure}

Figure~\ref{fig:one_bin} shows the four-fold differential cross sections as a function of $\phi$ for one representative bin in this newly measured region, with statistical and systematic uncertainties.    Cross sections predicted by BH and the KM15 \cite{Kumericki:2015lhb, Kumericki:2016ehc} and VGG \cite{Vanderhaeghen:1999xj} models are shown in blue, orange, and red, respectively.

The discriminating power of the data for different GPD models is quantified using the mean pull of the data with respect to the model normalized by the combined statistical and systematic (point-to-point and overall normalization) uncertainties.
The pulls with respect to KM15, shown in the lower panel of Fig.~\ref{fig:one_bin}, are distributed symmetrically around zero with a mean of 0.15, indicating that the KM15 model provides a consistent description of the cross section at these kinematics.
This suggests that KM15, a hybrid model fitted using the available DVCS data at both lower and higher $x_B$, interpolates reliably into this previously unexplored kinematic region.
In contrast, the mean pull with respect to VGG is -1.2.
The systematic offset of the VGG pulls demonstrates that the observed $\phi$-dependence 
is incompatible with models that significantly overestimate the cross section, confirming that these measurements provide meaningful constraints on GPD models in this previously unexplored kinematic region, and could serve as an input for future CFF extractions. 

\begin{figure*}[!ht]
  \includegraphics[width=0.9\textwidth]{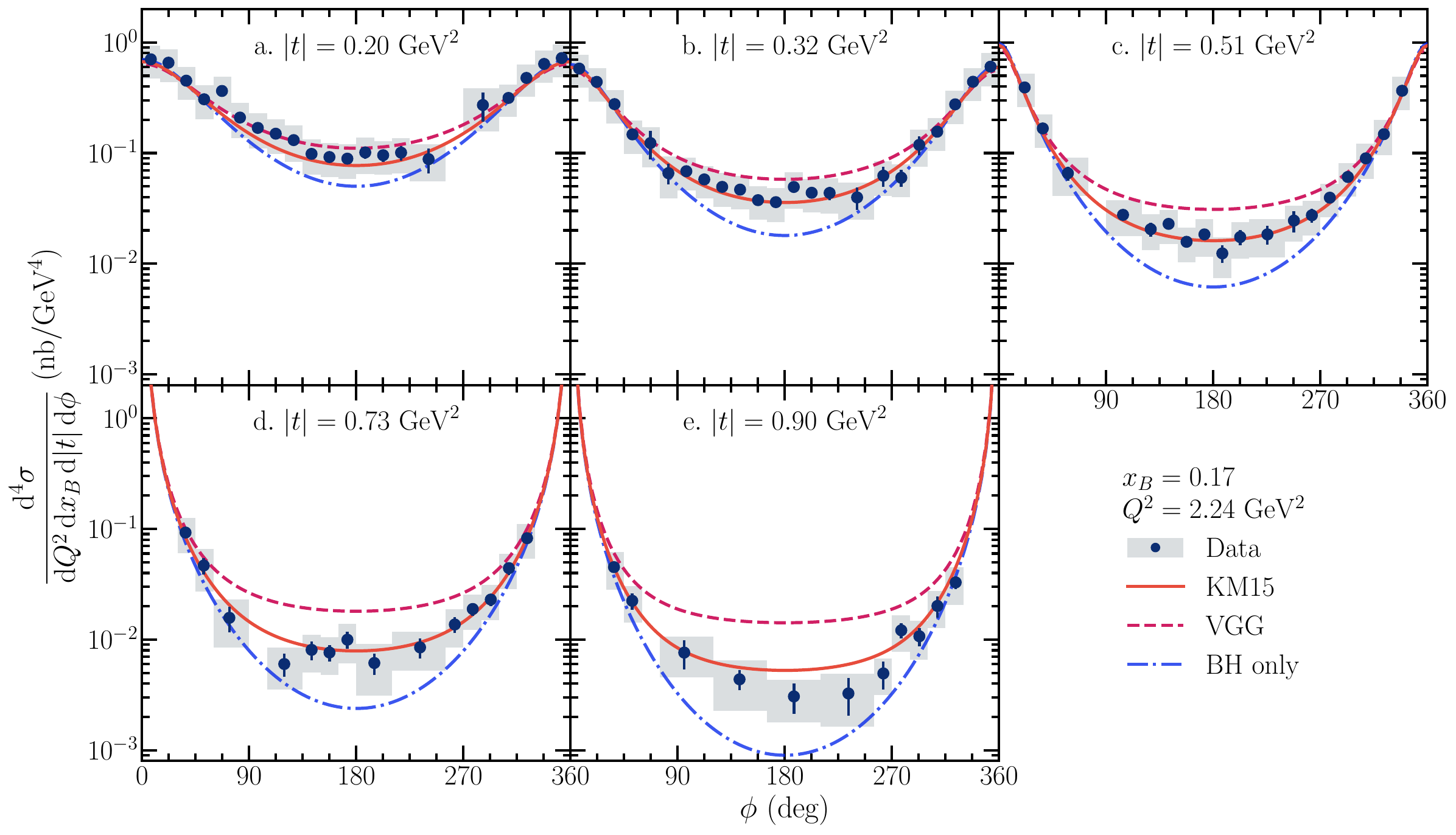}
    \caption{Unpolarized cross sections $\frac{d^4\sigma_{ep\rightarrow e'p'\gamma}}{dx_B dQ^2 d|t| d\phi}$ 
as a function of $\phi$ at $x_B = 0.17$ and $Q^2 = 2.24$~GeV$^2$ on average, 
for average $|t| = 0.20$ (a), 0.32 (b), 0.51 (c), 0.73 (d), and 0.90~GeV$^2$ (e).
Dark blue points represent the extracted cross sections with statistical (vertical bars) and systematic (gray bands) uncertainties. The blue dash-dotted, red dashed, and orange solid curves show the BH, VGG \cite{Vanderhaeghen:1999xj}, and KM15 \cite{Kumericki:2015lhb, Kumericki:2016ehc} predictions, respectively.}
    \label{fig:xsec_various_t}
\end{figure*}

\begin{figure*}[!ht]
    \includegraphics[width=0.9\textwidth]{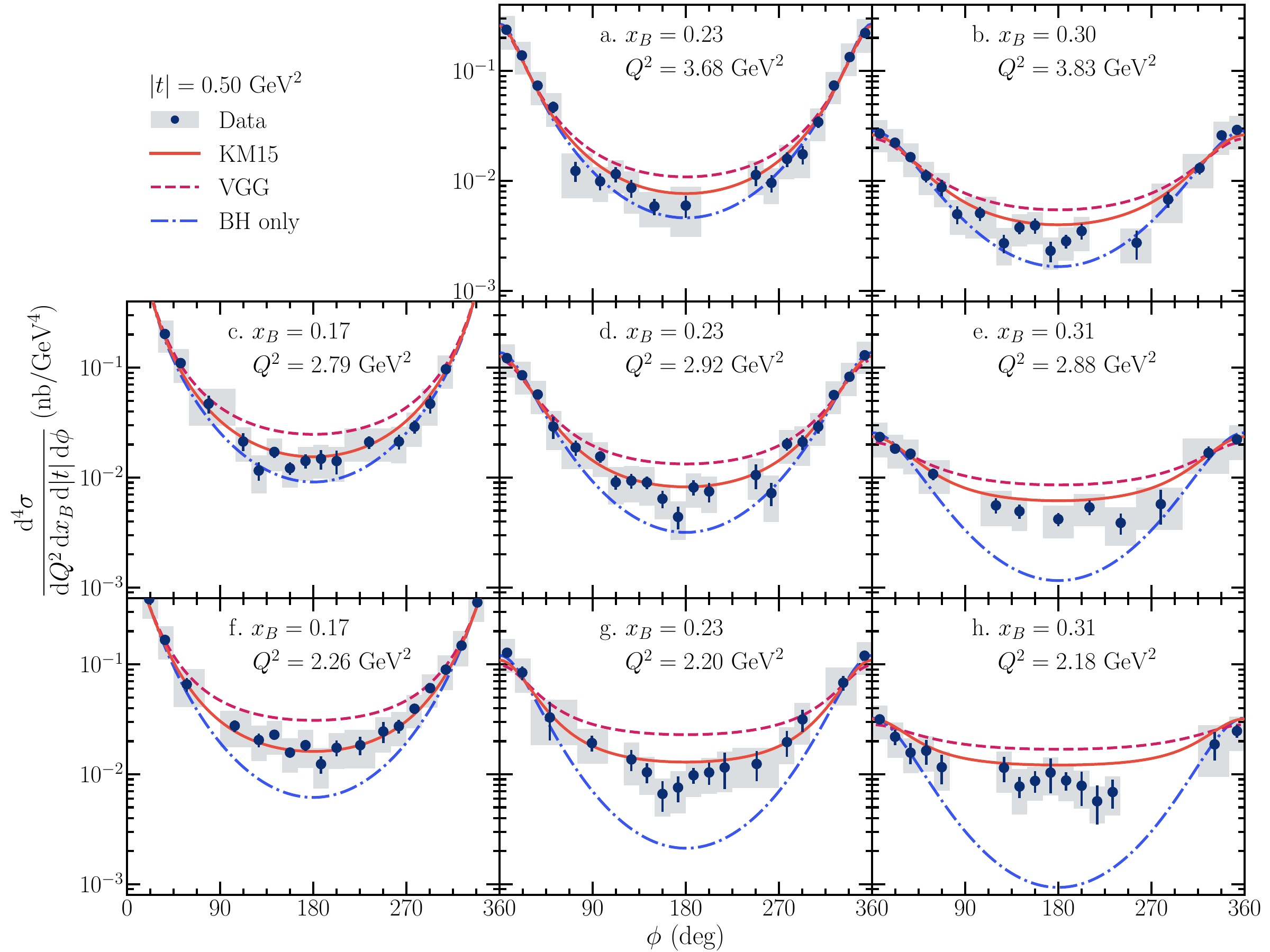}
    \caption{Unpolarized cross sections $\frac{d^4\sigma_{ep\rightarrow e'p'\gamma}}{dx_B dQ^2 d|t| d\phi}$ 
as a function of $\phi$ at average $|t| = 0.5$~GeV$^2$ for $0.16 < x_B < 0.36$ 
and $1.91 < Q^2 < 4.33$~GeV$^2$. Each panel shows the kinematics of a specific bin. Dark blue points represent measured cross sections, with statistical (vertical bars) and systematic (gray bands) uncertainties. The blue dash-dotted, red dashed, and orange solid curves show the BH, VGG \cite{Vanderhaeghen:1999xj}, and KM15 \cite{Kumericki:2015lhb, Kumericki:2016ehc} predictions, respectively.}
    \label{fig:xsec_various_xQ2}
\end{figure*}

Figure~\ref{fig:xsec_various_t} extends the data of Fig.~\ref{fig:one_bin} 
across the full range of average $|t|$
at the same average $x_B$ and $Q^2$, 
a kinematic region previously inaccessible to CLAS~\cite{PhysRevLett.115.212003} and 
not covered by the Hall A result at 10.6 GeV~\cite{JeffersonLabHallA:2022pnx}, which 
reaches down only to $x_B$ = 0.36. The mean pulls with respect to KM15 across 
the five $|t|$ bins range from $-0.51$ to $+0.40$, indicating a consistent 
description. The mean pulls with respect to VGG are negative in all five bins, with their absolute 
values increasing monotonically from 0.15 to 3.38 as the average $|t|$ increases 
from 0.20 to 0.90~GeV$^2$. 
This suggests that the $t$ profile of the VGG model becomes increasingly inconsistent 
with the data at larger momentum transfer. Figure~\ref{fig:xsec_various_xQ2} 
presents cross sections across a wide range of $(x_B, Q^2)$ bins at fixed 
average $ |t| $ of 0.50 GeV$^2$, where the mean VGG pulls are 
negative in every bin, consistent with the trend observed in 
Fig.~\ref{fig:xsec_various_t}. The systematic negativity of the VGG pulls across all 12 bins in both figures, a pattern that cannot be attributed to the overall normalization uncertainty, confirms that the data retain meaningful discriminating power between GPD models throughout the explored phase space. All measured cross section data points are provided in the CLAS Physics Database \cite{CLASDB}.

In conclusion, we report measurements of the multi-differential DVCS cross section at 10.6~GeV over an extensively expanded valence quark region in $x_B$, $Q^2$, and $t$ including numerous points in previously unreported kinematics. These expanded data provide important experimental input for future extractions of leading-twist CFFs by supplying constraints for their model-independent determination.
When combined with complementary observables, such as the beam-spin asymmetry \cite{CLAS:2022syx}, these measurements enable quantitative access to the proton’s GFFs, allowing a precise determination of its internal mechanical structure, including the spatial distributions of pressure and forces, as shown in Refs.~\cite{PhysRevC.92.055202, PhysRevLett.115.212003, JeffersonLabHallA:2022pnx}.

Complementary DVCS analysis from CLAS12 experiments at 6.4~GeV, 6.5~GeV, 7.5~GeV, 8.4~GeV and 10.2~GeV beam energies on unpolarized proton targets are under analysis, and beam spin asymmetries with this dataset \cite{CLAS:2022syx} and on unpolarized deuterium targets \cite{CLAS:2024qhy} have recently been published. Analyses of CLAS12 longitudinally polarized NH$_3$ and ND$_3$ target data for DVCS, which will provide access to target- and double-spin asymmetries, are in progress~\cite{Pilleux:2024zow}. For completeness, we note that the Neutral Particle Spectrometer \cite{Hamdi:2026qbb} in Hall~C will also probe the $Q^2$-dependence of CFFs~\cite{Rafael:2025skq}. 

We acknowledge the outstanding efforts of the staff of the Accelerator and the Physics Divisions at Jefferson Lab in making this experiment possible.
This material is based upon work supported by the U.S. Department of Energy, Office of Science, Office of Nuclear Physics under Contract No.~89243126CSC000213.
This work was supported in part by the U.S. Department of Energy, the National Science Foundation (NSF), and the Italian Istituto Nazionale di Fisica Nucleare (INFN), the French Centre National de la Recherche Scientifique (CNRS), the French Commissariat à l’Energie Atomique (CEA), the UK Science and Technology Facilities Council (STFC), the National Research Foundation of Korea (NRF), the Helmholtz-Forschungsakademie Hessen für FAIR (HFHF), the Chilean Agencia Nacional de Investigacion y Desarollo (ANID), the Scottish Universities Physics Alliance (SUPA), the Skobeltsyn Nuclear Physics Institute and Physics Department at the Lomonosov Moscow State University. This research was supported by Grant No.~2024018 from the United States–Israel Binational Science Foundation (BSF).
\bibliography{main}

\end{document}